\documentclass[aps,pre,twocolumn,showpacs,groupedaddress,footinbib]{revtex4-1}
\usepackage{natbib,hyperref}
\hypersetup{colorlinks=true, citecolor=blue, urlcolor=blue, linkcolor=blue}
\usepackage[english]{babel}
\usepackage{graphicx}
\usepackage[caption=false]{subfig}
\usepackage{amsmath}
\usepackage{amsfonts}
\usepackage{amssymb}
\usepackage{color,soul}
\setulcolor{red}
\usepackage{xcolor}
\usepackage[utf8]{inputenc}
\begin{document}

\title{Phase separation in a binary mixture of sticky spheres}

\author{D. C. Thakur}
\author{Jalim Singh\textsuperscript{$\dagger$}}
\author{A. V. Anil Kumar}
\email{anil@niser.ac.in}
\affiliation{School of Physical Sciences, 
National Institute of Science Education and Research,
HBNI, Jatni, Bhubaneswar 752050, India.}
\affiliation{\textsuperscript{$\dagger$}Current Adress: Department of Biomedical Engineering, Oregon Health and Science University, Portland, OR 97239, U.S.A.}

\begin{abstract}
We numerically investigate the dependence of range of attractive potential on the phase separation of 2-D binary systems. Through extensive simulations and analysis, we show that when the range of attractive interactions approaches the sticky sphere limit, the system undergoes a phase separation at lower temperature. Further reduction in temperature causes the system to mix again. These mixing-demixing-mixing transitions are of first order. Such phase separation is not observed for systems with larger interaction range. In the phase separated region of the phase diagram, one of the components of the mixture chooses to be in crystalline configuration, while other being in disordered state. 
\end{abstract}

\maketitle

\section{Introduction}
Phase separation in binary mixtures is a complex and multifaceted phenomenon that has garnered substantial attention across a spectrum of scientific disciplines\cite{j:erin_2024,j:pritha_2018,j:christoph_2019,j:meer_2016,j:michio_2021}. This intricate process involves the demixing of components, resulting in the emergence of distinct phases \cite{j:maity_2023,j:shan_2023,j:erin_2024}. Colloidal systems have been at the forefront of research in understanding phase separation, particularly within the realm of soft matter physics. Studies on colloidal glasses by the scientific community and investigations into ordered structures in binary mixtures  have played pivotal roles in unraveling the nature and kinetics of phase separation \cite{j:nele_2023,j:lijie_2022,j:joep_2020,j:tata_2006,j:pedersen_2018,j:jalim_2020,j:tanaka_2022}. The interplay of attractive and repulsive forces in colloidal systems, those modeled by the different potential functions, significantly influences the resulting phase behavior \cite{j:franco_rev_2021,j:marques_2015,j:louis_2019,j:marco_2016,j:likos_2007,j:brian_2022,j:anna_2020}. Furthermore, the exploration of liquid-liquid phase separation \cite{j:mehzabin_2023,j:ahmed_2023,j:lev_1996} and gas-liquid phase separation \cite{j:ryuichi_2016,j:jack_2024,j:pablo_2019} has added insights, enriching our understanding of the diverse manifestations of phase separation in binary mixtures.

Phase separation in binary mixtures is usually believed to be induced by attractive interactions between the particles rather than the short ranged repulsive interactions. Later on it is revealed that phase separation can occur in systems with purely repulsive interactions\cite{j:kerley_1989,j:kujiper_1990,j:biben_1991}. Effect of these short range repulsive interactions as well as the attractive interactions has been studied extensively\cite{j:kambayashi_1992,j:pellicane_2006, j:alonso_1997,j:foffi_2005}. For purely repulsive inverse power potential, the phase separation depends on the softness of the repulsive potential\cite{j:kambayashi_1992}. In binary mixtures with attractive interactions, the phase separation at lower temperatures is often interrupted by an arrested glassy phase\cite{j:foffi_2005}. The studies on the interplay between the attractive and repulsive forces has provided valuable insights into the thermodynamics and kinetics of phase separation in binary mixtures \cite{j:hoang_2022,j:kajetan_2016,j:umberto_2011,j:rajamani_2019}. 

In colloidal systems, it is possible to realise a system of interacting particles whose interaction can be tuned by various means; i.e. interactions can be controlled in strength as well as range\cite{b:russel}. 
The exploration of phase separation in binary colloidal mixtures and its dependence on attractive interactions are mainly confined to the strength of interactions. There are fewer studies about the  dependence of range of attractive interactions on the stability of binary mixtures. The main impedence on this front was the nonexistence of a potential function whose attractive range can be varied. Recently, Wang {\it et al.}\cite{j:rc_pot} proposed an interparticle interaction whose range of attractive well can be varied. At larger ranges, the potential is very similar to Lennard-Jones potential and exhibits a similar phase diagram. At very low ranges, the potential reduces to sticky spheres. In this article, we make use of this potential to investigate the effect of the range of attractive potential on the stability of binary mixtures using Langevin dynamics simulations. We found that while approaching the sticky sphere limit, the binary mixture spontaneously phase separates into large particle-rich and small particle-rich phases.  

The remainder of this article is organised as follows. In section II, we detail our model and the interaction potentials. The simulation details are also outlined in this section. Our results are presented in section III along with discussions. We finally summerise the main results in section IV.

\section{Modeling and simulation framework}

We performed Langevin dynamics simulations of a binary mixture in two dimensions. The binary system studied here  composed of 1000 particles with a size ratio  1:1.4 and the composition ratio  50:50\cite{j:onuki_2dglass,j:tanaka_2dglasorder}. Hereafter, the larger particles are referred as A particles and smaller particles as B particles. The components in the mixture are interacting via the potential
function given by equation \ref{e:pe1}\cite{j:rc_pot}. 
\begin{equation}
\label{e:pe1}
\phi(r_{ij}) = \begin{cases}
\epsilon\zeta \left(\left[\frac{\sigma_{\alpha\beta}}{r_{ij}}\right]^{2}-1\right) \left(\left[\frac{r_c}{r_{ij}}\right]^{2}-1\right)^{2} & \text{for } r_{ij}\leq r_c, \\
0 & \text{for } r_{ij} > r_c.
\end{cases}
\end{equation}

This potential is special case of the general form of potential proposed by Wang {\it et al.}\cite{j:rc_pot},

\begin{equation}
	\label{e:pe}
	\phi(r_{ij}) = \epsilon\zeta \left(\left[\frac{\sigma_{\alpha\beta}}{r_{ij}}\right]^{2\mu}-1\right) \left(\left[\frac{r_c}{r_{ij}}\right]^{2\nu}-1\right)^{2\nu}
\end{equation}

where the coefficient $\zeta$ ensures that the depth of the attractive well is -$\epsilon$, while $\mu$, and $\nu$ are all positive integers. This potential has been proposed as an alternative to Lennard-Jones potential when the interactions in a system are finite ranged. Here the cut-off parameter $r_c$ determines the range of interactions and the potential vanishes quadratically at $r_c$.  We consider the case ($\mu = \nu = 1$) in equatin \ref{e:pe}, for which potential has the form given by equation \ref{e:pe1}. With this choice of $\mu$ and $\nu$ and the constraint that the value of the potentail at the minimum is -$\epsilon$, we get that 

\begin{equation}
	\zeta(1,1;r_c) = 2\left(\frac{r_c}{\sigma_{\alpha\beta}}\right)^2\left(\frac{3}{2\left(\left(\frac{rc}{\sigma_{\alpha\beta}}\right)^{2}-1\right)}\right)
\end{equation}

and

\begin{equation}
	r_{min}(1,1;r_c) = r_{c}\left(\frac{3}{1+2\left(\frac{r_c}{\sigma_{\alpha\beta}}\right)^2}\right)^{1/2}
\end{equation}

\begin{figure}
	\includegraphics[width=8.5cm, height=6.5cm]{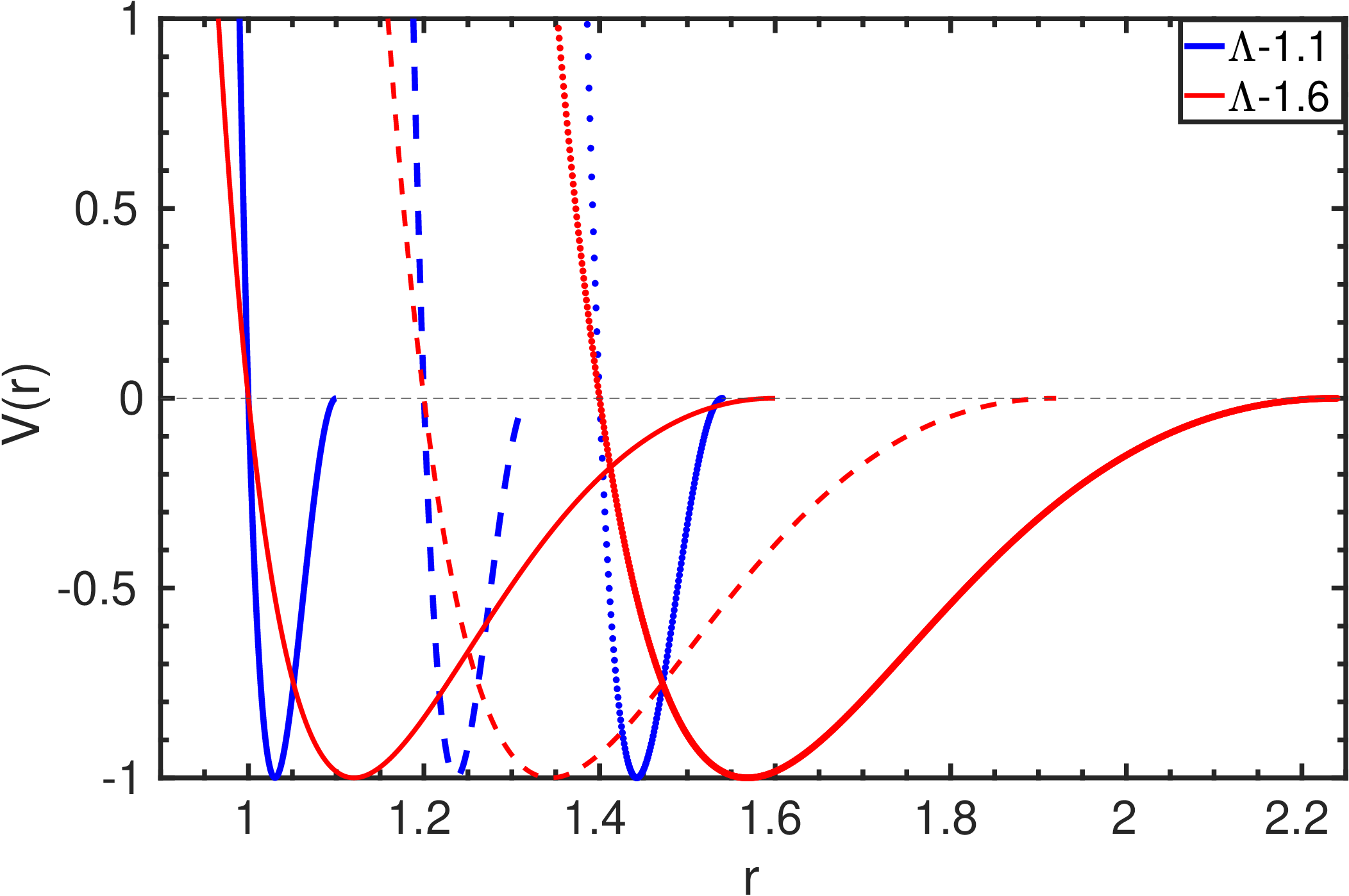}
	\caption{\label{f:poten} Potential used in the study is plotted separately for AA (-- lines), AB (- - lines) and BB (... lines) interactions, at $\Lambda=$1.1 (lowest simulated $\Lambda$) and 1.6 (largest simulated $\Lambda$). For example for $\Lambda=$1.6 in case of AA prticles the potential is zero beyond $r=$2.24.}
\end{figure}

One advantage of using this potential is that we can set the range of potential by setting the value of $r_c$ in \ref{e:pe1}. We have employed this potential in our simulations with $r_c$ defined as $r_c=\sigma_{\alpha\beta} \times \Lambda$ for different pair interactions, where $(\alpha, \beta) \in (A, B)$. Here $\Lambda$ is the range of attractive interaction. We have varied the $\Lambda$ ranging from 1.1 to 1.6 for the pair interactions. Figure \ref{f:poten} depicts the A-A, B-B and A-B interactions for two different ranges.

The dynamics of the particles in the binary system is governed by the Langevin's equations of motion (see eq. \ref{e:lde}).
\begin{equation}
	\label{e:lde}
	m_i \ddot{\mathbf r_i} = -\gamma \dot{\mathbf r_i} + \sum_{ij}
	\mathbf F_{ij} + {\sqrt{2k_BT\gamma}} {\boldsymbol {\eta}_i} , 
\end{equation}
Here $\gamma$ denotes the friction coefficient, 
and the interparticle interaction force was calculated as $\mathbf F_{ij}=- \boldsymbol{\nabla} V(r_{ij})$ , T denotes the thermodynamic temperature at which the simulation is performed, and $\eta_i$ denotes the random gaussian noise with zero mean and  unit variance, \textit{i.e.}, $\langle \eta_i \rangle =$0 and 
$\langle \eta_{i\alpha}(t)\eta_{j\beta}(t^{\prime}) \rangle = 
2k_B T\gamma \delta_{ij}\delta_{\alpha\beta}\delta(t-t^{\prime})$.

The simulation was performed at a constant area fraction $\phi=$ 0.628, calculated as $\phi=\pi\rho(\sigma^2_{AA}+\sigma^2_{BB})/8$. For the underdamped condition, the damping coefficient $\gamma$ is set to 10.0. Equations of motion were integrated with a time step $dt=$ 0.002 using the algorithm given in \cite{j:eijden-ciccotti}. All quantities are presented in Lennard-Jones (reduced) units, such that reduced density $\rho^*=\rho\sigma^3$, reduced temperature $T^*=k_BT/\epsilon$, reduced time $t^*=(\epsilon/{m\sigma^2})^{1/2}t$, and reduced force $f^*=f\sigma/\epsilon$ \cite{ham}.
\begin{figure}
	\includegraphics[width=8.5cm, height=8.5cm]{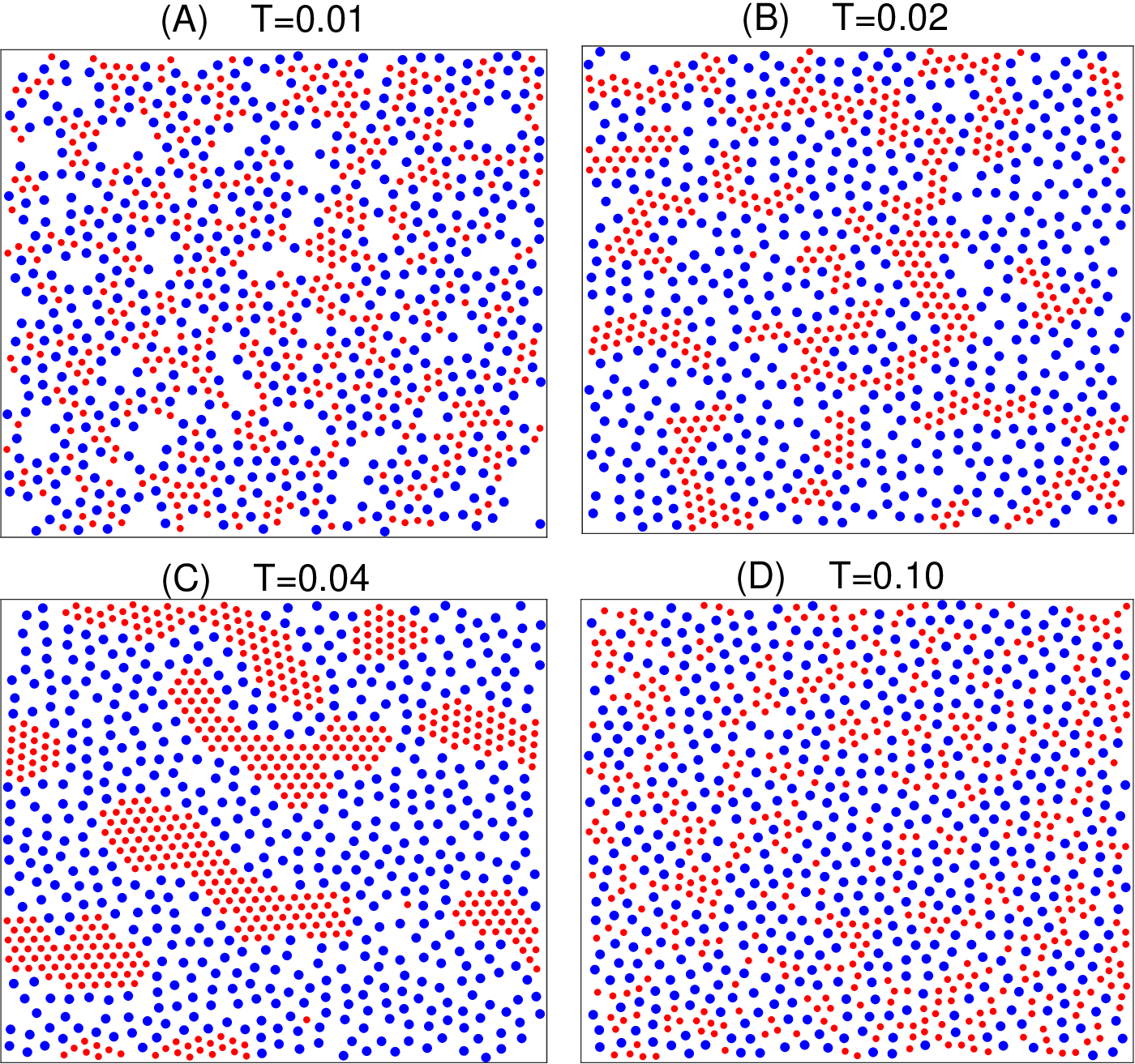}
	\caption{\label{f:cnf} Figure (A), (B), (C), and (D) shows the final configurations of the simulation at $\Lambda=$1.1 for temperatures $T=$ 0.01, 0.02, 0.04 and 0.10 respectively. The blue and red colour represent A (large) and B (small) particles respectively.}
\end{figure}

A random initial configuration was produced at an area fraction $\phi=$ 0.628 for each $\Lambda$ at high temperature $T$ = 8.0 and for simulations at this high temperature, the time step of integration was kept $dt=$0.001. The final configuration obtained from these high-temperature simulations was rapidly cooled to corresponding lower temperatures, equilibrated for $7.5\times 10^7$ steps at all temperatures and $\Lambda$, and subsequently, trajectories were stored for further analysis.

\section{Results}
\begin{figure}
	\includegraphics[width=8.5cm, height=7.5cm]{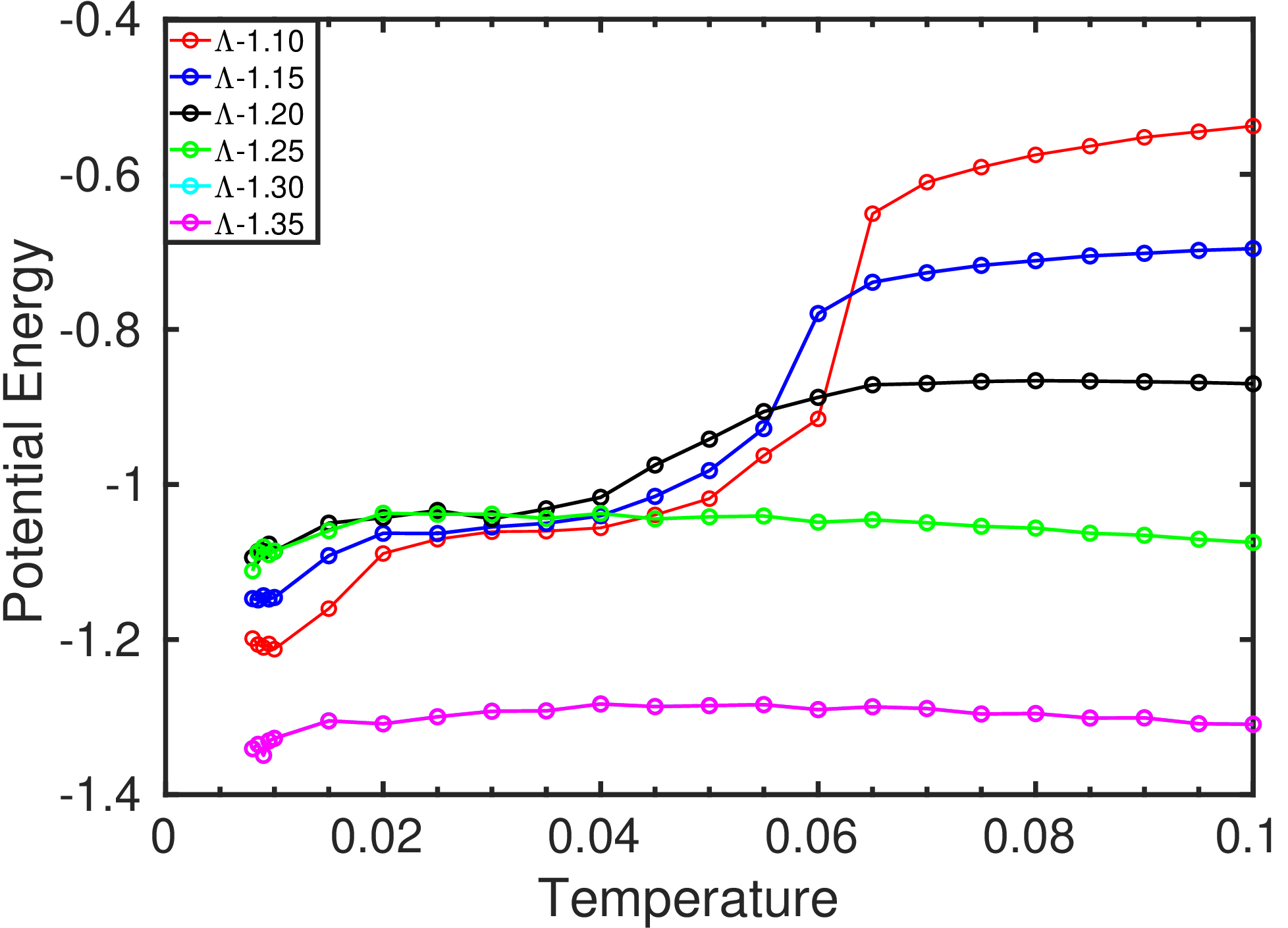}
	\caption{\label{f:pete} Potential energy plotted at different $\Lambda$ as a function of temperatures.}
\end{figure}

We have carried out Langevin dynamics simulations for different values of $\Lambda$ as well as at different temperatures. 
The final configuration obtained for  $\Lambda=$1.10 at different temperatures is shown in figure \ref{f:cnf}. At very low temperatures, as illustrated in figure \ref{f:cnf}(A) for $T=$0.01, the system is in a mixed state. As we increase the temperature to $T=$0.02 (see figure \ref{f:cnf}(B)), the system start demixing and the B (red)  particles start forming small domains. On further increasing the temperature the domain size of the cluster of B particle increases (as depicted in figure \ref{f:cnf}(C) for $T=$0.04). On increasing the temperature further, the system undergoes mixing again, as shown in \ref{f:cnf}(D) for $T=$0.1. Similar behaviour was observed for $\Lambda=$1.15 and 1.20. However this mixing-demixing-mixing behaviour has not been observed for higher values values of $\Lambda$ and the system is always in a mixed state. In figure \ref{f:pete}, we have plotted the potential energy of the system against temperature for different values of $\Lambda$. 
It is clear from the figure that for $\Lambda = $ 1.10, 1.15 and 1.20, there is abrupt increase in the potential energy curve corresponding to mixing-demixing transition and demixing-mixing transition, confirming these transitions are of first order in nature. However, for higher values of $\Lambda$ where we do not see any transitions, the potential energy curve is continuous.

\begin{figure}
	\includegraphics[width=8.5cm, height=8.5cm]{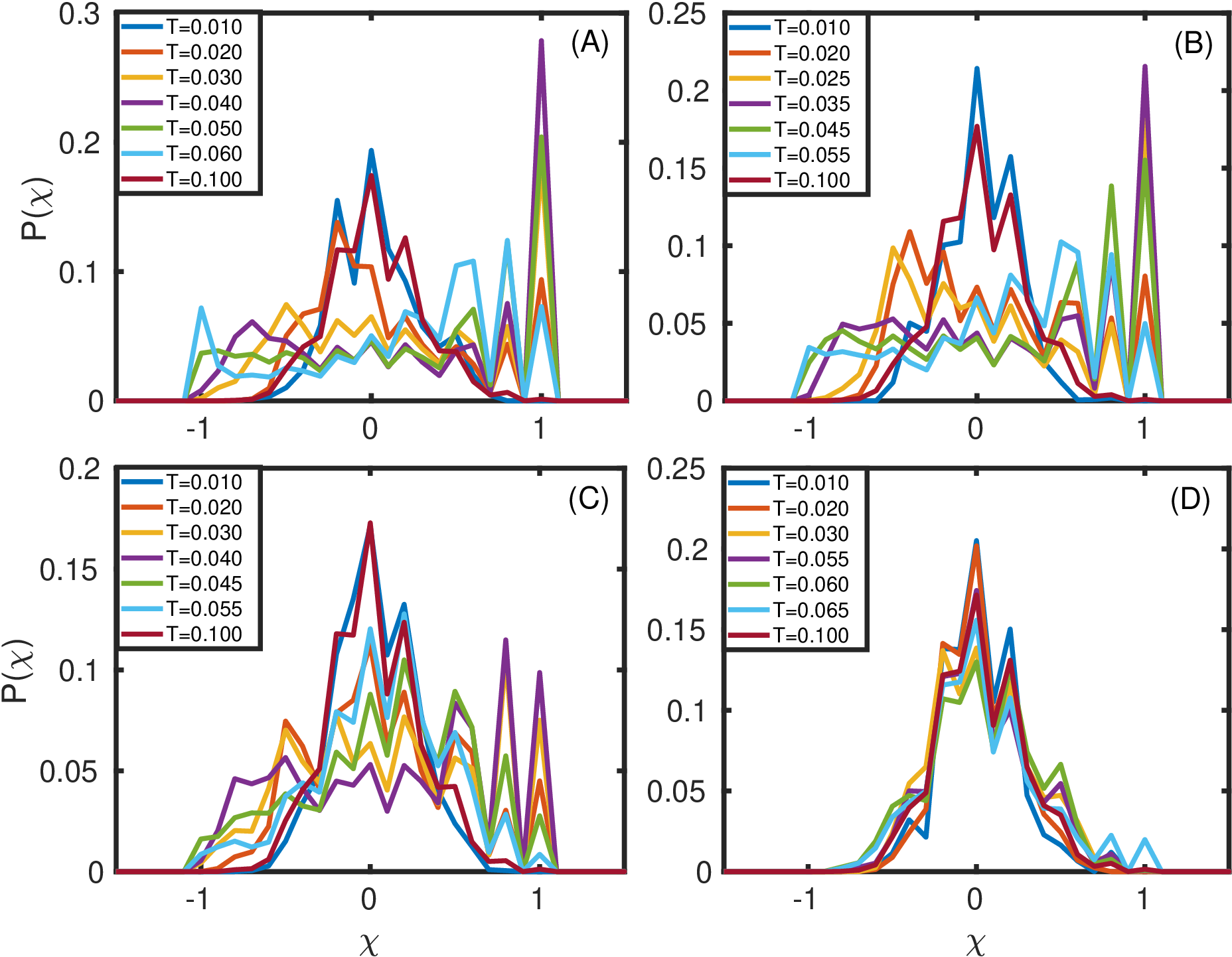}
	\caption{\label{f:kipki} The $\chi$ distribution, denoted as $P(\chi)$, depicted in figure (A), (B), (C), and (D) at $\Lambda=$1.10, 1.15, 1.20 and 1.25 respectively, at temperatures shown in the legend.}
\end{figure}


In order to quantify this phase separation, we have divided the simulation box into a number of square subcells($N_{cell}=8\times8$). Each of this subcells has a cell length of   $l_c=5.38$. For a given equilibrium configuration, we calculated the difference in number  density of A and B type particle as $\chi=(n_A^i-n_B^i)/(n_A^i+n_B^i)$\cite{j:activ_phassep_chandan}. Here, $n_A^i$ and $n_B^i$ denote the counts of $A$ and $B$ particles, respectively, in the $i^{th.}$ cell. Absolute value of this quantity is denoted as the order parameter for this mixing-demixing transition. This is be averaged over all the subcells as well as many equilibrium configurations and 
is expressed as:

\begin{equation}
\label{e:psop}
\Phi(\Lambda,T) = \frac{1}{N{cell}} \left\langle \sum_{i=1}^{N_{cell}}
\frac{|n_A^i-n_B^i|}{(n_A^i+n_B^i)} \right\rangle,
\end{equation}

\begin{figure}
	\includegraphics[width=8.5cm, height=7.0cm]{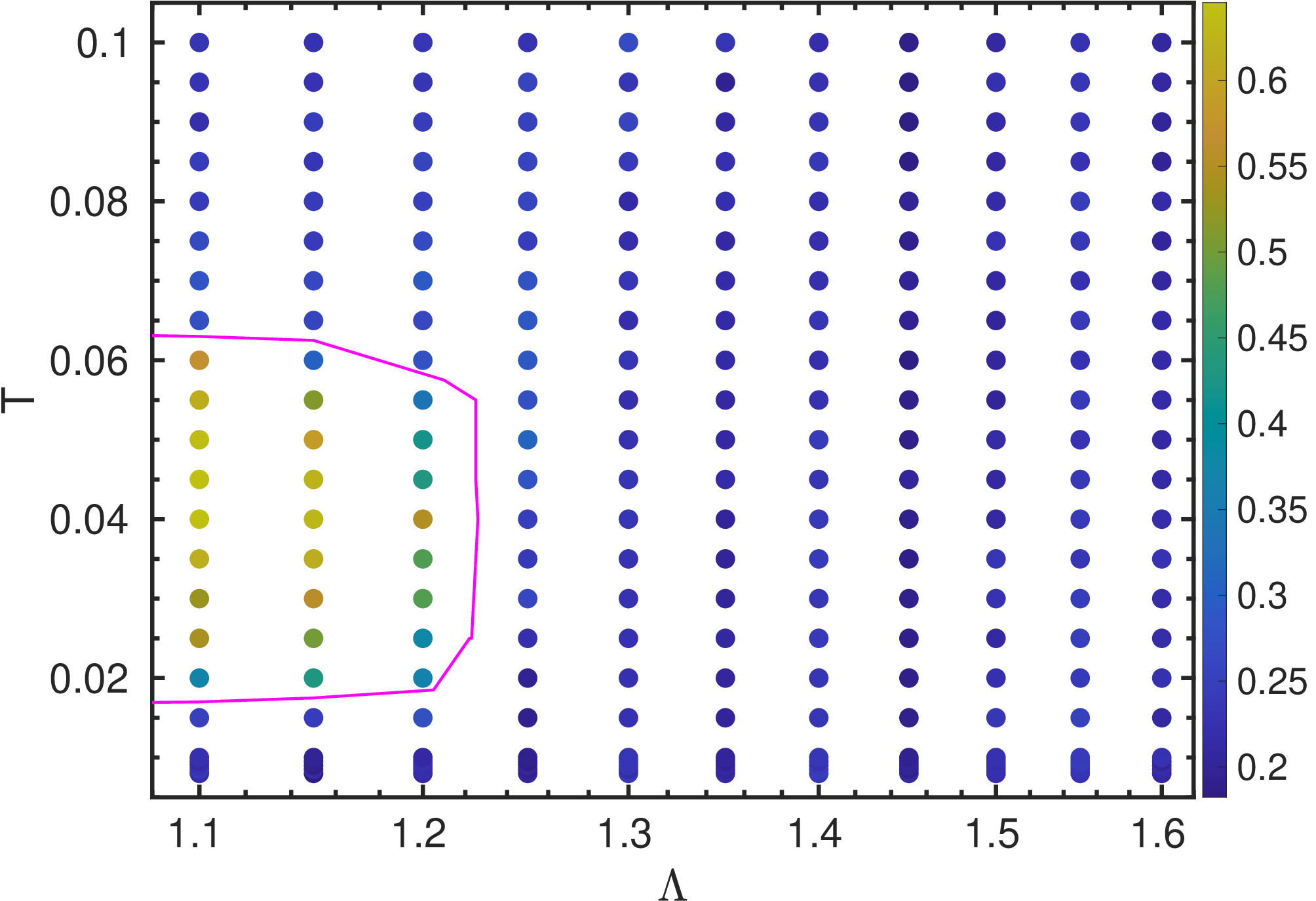}
	\caption{\label{f:phase} Phase diagram at different $\Lambda$ and temperatures, the color bar corresponds to the average value of $\chi$, the lightest yellow colour corresponds to highly demixed state while the darkest blue colour represents highly mixed state.}
\end{figure}

Within Equation \ref{e:psop}, the angular brackets indicate averaging over the system's equilibrium configurations. This order parameter $\Phi(\Lambda,T)$ acts as a metric for gauging whether the system is in a mixed or demixed state. If $\Phi(\Lambda,T)$ is below its critical threshold, the system is deemed mixed; conversely, if $\Phi(\Lambda,T)$ exceeds its critical threshold, the system is considered demixed. The analysis of onset of phase separation in the system focuses on a critical value of $\Phi(\Lambda,T)$, termed as $\Phi_c(\Lambda,T)$ for this study. 

To calculate $\Phi_c(\Lambda,T)$,  we have calculated  $\chi$  for the subcells  over the  equilibrium configurations and formed the distribution of $\chi$, denoted as $P(\chi)$. This is done for   different values of $\Lambda$ as well as temperatures.  These distributions, $P(\chi)$, are plotted in  figure \ref{f:kipki}(A), (B), (C) and (D) for $\Lambda=$1.10, 1.15, 1.20 and 1.25 respectively, at various temperatures. Figure \ref{f:kipki}(A) illustrates that $P(\chi)$ shows a prominent peak near $\chi=0$ at $T=0.01$, which indicates that the most of the subcells have equal number of A and B type of particles and the system is in a mixed state. As we increase the temperature,  the height of  this peak starts diminishing accompanied by broadening. Further increase in temperature results in splitting of this peak into two at non-zero values,  indicating a disparity in the number of A and B particles in the subcells. The appearance of this two peak structure thus indicates the onset of phase separation of the binary mixture.   The broadening and appearance of double peak structure in $P(\chi)$ starts at $T=0.02$  and we designate the value of $\Phi(\Lambda,T)$ at this temperature as the threshold $\Phi_c(\Lambda,T)$, indicating the initiation of the phase separation. With further increase in temperature,
both peaks grow, indicating that the demixing becomes stronger. Please note that the  peak at $\chi= \pm$1 becomes more prominent as we increase the temperature, which shows that the two phases in the  demixed system are prominently made up of A or B particles.   Figure \ref{f:kipki}(A) reveals that, beyond temperature $T=$0.06, the two peaks start coming closer and merges, leading to one peak at $\chi$=0 at $T=0.1$, signifying that the system prefers to be in a mixed state at higher temperature.  This is evident  in the snapshot, plotted in figure \ref{f:cnf}(D)  for temperature $T=$0.1.  Similar behaviour in $p(\chi)$ observed for  $\Lambda=$1.15 and 1.20, however the extent of  phase separation lessens owing to the increase in the softness of the potential, resulting in a decline in the maximum peak height observed in $P(\chi)$.  For $\Lambda=$ 1.25 and above, we do not observe any double peak structure in $P(\chi)$, suggesting  that no mixing-demixing-mixing transition occurs if the range of attractive potential is larger.  Thus , the demixed state observe at these lower temperatures occurs only for sticky spheres, where the range of attractive interaction is very small compared to the size of the particles.  In Figure \ref{f:phase}, we present the phase diagram in the $\Lambda-T$ plane, where the color bar corresponds to the average value of $\chi$, $\Phi(\Lambda,T)$.  We have plotted the value of  $\Phi(\Lambda,T)$ for all 253 state points simulated the colour in phase diagram.  Demixed states are represented by the color on the top side of the color bar, while the mixed states are represented by the bottom side. We used the value of $\Phi_c(\Lambda,T)$, as described earlier to determine the phase boundary of demixed and mixed state.

\begin{figure}
	\includegraphics[width=9.0cm, height=7.0cm]{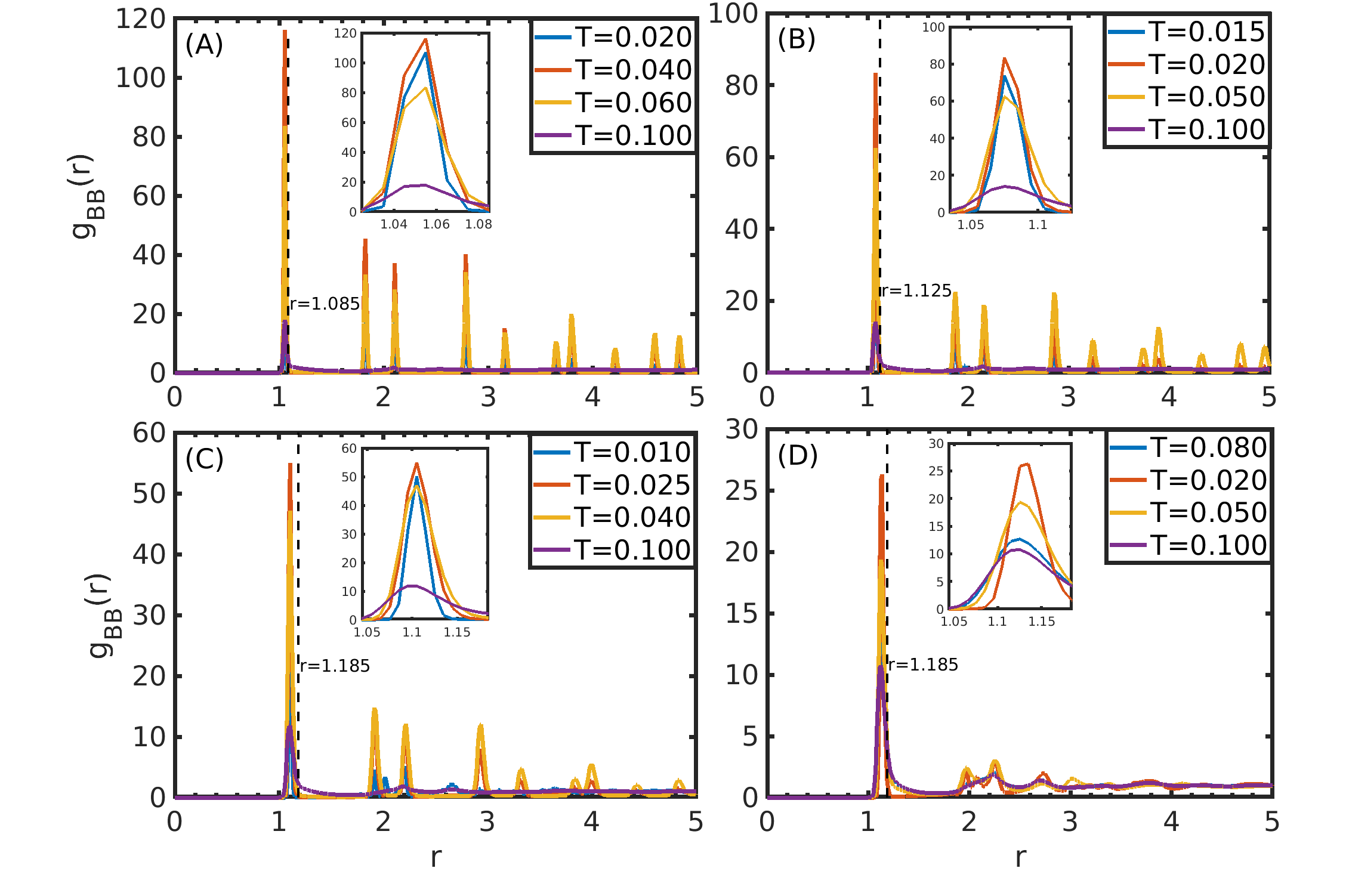}
	\caption{\label{f:grbb} The Radial Distribution Function $g_{BB}(r)$ depicted in figure (A), (B), (C), and (D) at $\Lambda=$1.10, 1.15, 1.20 and 1.25 respectively, at temperatures shown in the legend.}
\end{figure}
The mixing and demixing of the particles with the temperature variation at different values of $\Lambda$, is accompanied by structural changes in the system as suggested by the instantaneous configurations plotted in figure  \ref{f:cnf}.  To investigate these structural changes in the system in mixed and phase seprated states, we first calculated calculate the  radial distribution function(RDF). As evident from the snapshot configurations, one of the separated phases is dominated by the smaller B particles, which appears to form hexagonal crystallites.  So we have calculated   $g_{BB}(r)$, which is shown in figure \ref{f:grbb} (A), (B), (C) and (D) for $\Lambda=$1.10, 1.15, 1.20 and 1.25 respectively at few representative temperatures. For $\Lambda=$ 1.10 the intensity of the first peak of RDF increases from temperature $T=$0.02 to $T=$ 0.04, indicating the accumulation of the more B neighbours around B with increasing temperature.  There are multiple sharp peaks at larger values of $r$, signifying long range ordering in the system.  These peak positions are matching with two dimensional hexagonal crystalline phase. On increasing the temperature further, the intensity of these peaks  reduces. Thus we can see that the long range ordering in the system initially increases with the temperature and then starts decreasing with the further temperature increment. For this particular $\Lambda$, we can see that the ordering is less in the system for lower and higher temperature side, as compared to the intermediate temperatures, indicating that the mixing-demixing-mixing transitions are accompanied by the formation and decomposition of hexagonal clusters of B particles and similar behaviour is observed in the $g_{BB}(r)$ for $\Lambda=$1.15 and 1.20, where we observed demixing at intermediate temperatures. But, for $\Lambda=$1.25 and higher values of $\Lambda$ studied (figure \ref{f:grbb} (D)),  we do not find any long range ordering in the $g_{BB}(r)$ at any temperature. 

\begin{figure}
	\includegraphics[width=8.5cm, height=7.5cm]{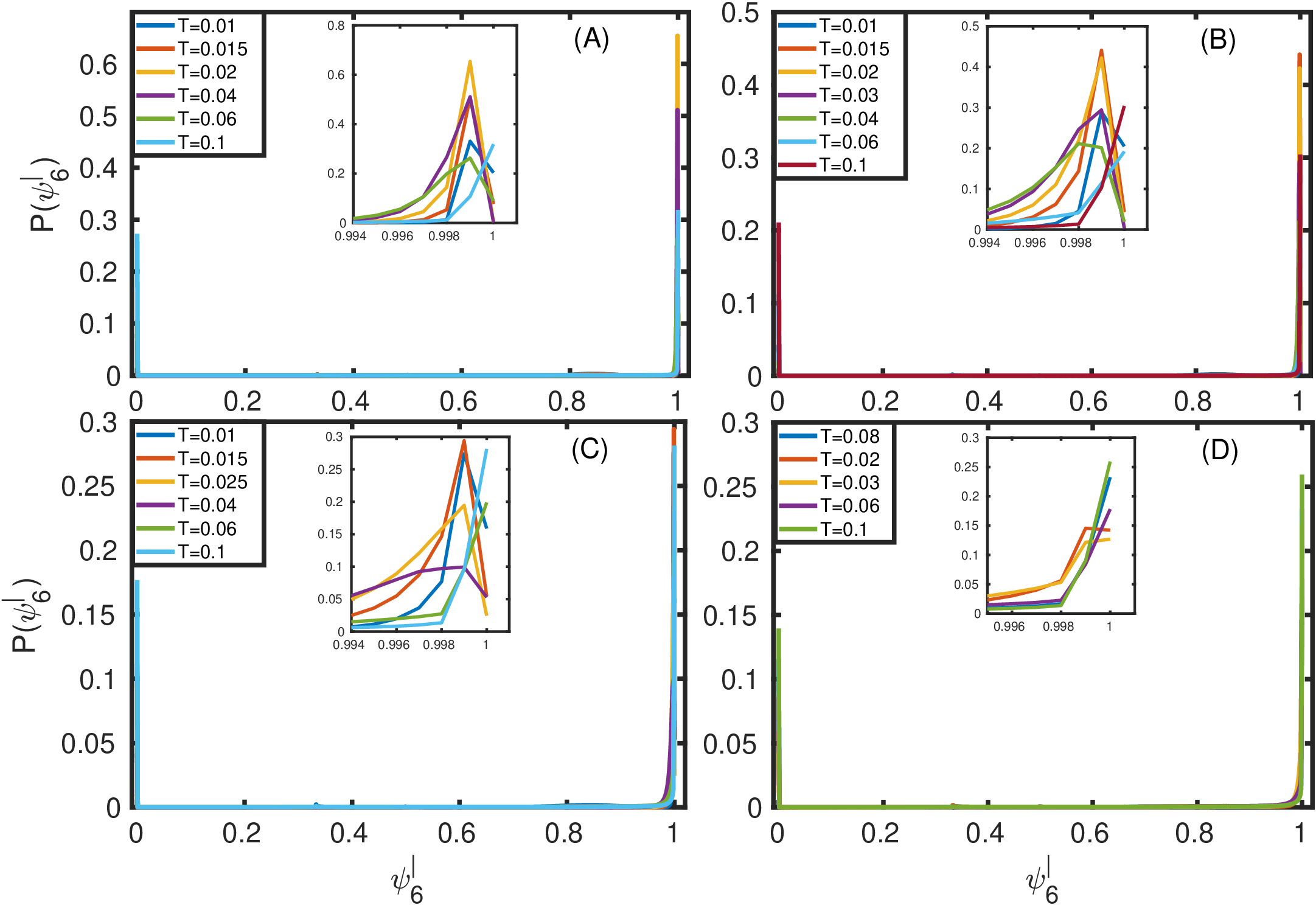}
	\caption{\label{f:hopbb} The Hexatic Order Parameter depicted in figure (A), (B), (C), and (D) at $\Lambda=$1.10, 1.15, 1.20 and 1.25 respectively, at temperatures shown in the legend.}
\end{figure}

The radial distribution function of smaller B particles, where the system is in a demixed state, we have seen that there is a long range order and the peaks corresponds to hexagonal closed packed structure. In order to further validate this, we have calculated the orientational ordering of the B particles. Utilizing the single-particle Hexatic Order Parameter (HOP), we investigate the local orientational ordering within the binary system. The hexatic order parameter is defined as:

\begin{equation}
\label{e:psi6}
\psi_6^j = \frac{1}{N_b^j} \sum_{m=1}^{N_b^j} \exp(\imath 6\theta_j^m),
\end{equation}
where $N_b^j$ represents the count of nearest neighbours surrounding particle $j$, and $\theta_j^m$ measure the angle subtended by $\mathbf{r}_j^m$ on the reference axis, where $\mathbf{r}_j^m$ is the radius vector. \cite{j:hardisk_KTHNY,j:onuki_2dglass}. Depending on $\theta_j^m$, the magnitude of $|\psi_6^j|$ varies between 0 and 1; a perfect trigonal configuration corresponds to $6\theta_j^m=2\pi$, resulting in $|\psi_6^j|=1$. while $|\psi_6^j|=0$, it indicates a random orientational arrangement of the six nearest neighbors of particle $j$. To determine the cut-off for computing the HOP, we analyze the radial distribution function (RDF) for B particles, singling out the end point of the first maximum as the cut-off range, representing the range of the first coordination shell (see black dashed line in figure \ref{f:grbb}). As indicated by the RDF and $\Phi(\Lambda,T)$, the local ordering of the B particles cause the phase separation in the system. Hence, we analyze the averaged distribution of local hexatic ordering across the steady-state configurations of the system. Figure \ref{f:hopbb} represents the distribution of the single-particle hexatic order parameter for B particles, labeled as $P(\psi_6^l)$,  for $\Lambda=$1.10, 1.15, 1.20 and 1.25 respectively. Figure \ref{f:hopbb}(a) demonstrates that $P(\psi_6^l)$ features a peak near $\psi^l_6\simeq$ 1. This peak indicates the local hexagonal ordering. For very small temperature, this peak is not significant. However, as we increase the temperature, the peak at  $\psi^l_6\simeq$ 1 grows indicating increase in local hexatic arrangement.  The highest peak in $P(\psi_6^l)$ occurs at $T=0.02$, where the binary mixture starts the phase separation. Beyond $T=0.02$ for $\Lambda=$1.10, the peak of $P(\psi_6^l)$ near $\psi^l_6\simeq$ 1 begins to decrease, representing smaller values of $P(\psi_6^l)$ compared to those at $T=0.02$.  This indicates that at this particular temperature the orientational ordering is highest as compared to other temperatures. Further increment in the temperature, reduces the intensity of the peak, suggesting that temperature elevation is affecting the orientational ordering in the system. As evident from the final configuration of the system (Figure \ref{f:cnf}) at various temperatures, with the increase in the temperature the domain size of  phase separated B particle also increases initially. This decreases as system start mixing again. The largest peak at $\chi=$1 indicates that at this particular temperature one type of particles are aggregated more in different regions in the box. This is supported by the $g_{BB}(r)$, where the largest height in the first peak indicate that there are more B particles around any B particle.  For $\Lambda=$1.15 and 1.20 (see figure \ref{f:hopbb}(B) and (C)), $P(\psi_6^l)$ exhibits qualitative similar behaviour as observed at $\Lambda=$1.10, but with a smaller extent of orientational ordering in B particles, as indicated by the intensity of the peak at $\chi=$1.  But such hexagonal ordering decreases for higher values of $\Lambda$ and we do not observe any increase in hexatic order parameter for intermediate temperature in those cases.

After analyzing the instanteneous configuration of the simulation, radial distribution function, $\Phi(\Lambda,T)$, and HOP for various $\Lambda$, we can come to a conclusion that phase separation coincides with hexatic ordering in B particles.  This phase separation and associated hexagonal ordering occurs in the sticky sphere limit of attractive interactions. We do not find any evidence of such a mixing-demixing-mixing transition for larger values of $\Lambda$. This phase separation is found to be a first order transition as indicated by the potential energy versus temperature curve. However, it should be noted that the second phase in the phase separated system which is dominated by the larger particles is not ordered and appears to be in a glassy state.

\section{Conclusion\label{s:cncl}}
Based on our numerical investigations into the phase behaviour of 2-D binary systems, we have uncovered several insights into the dependence of the range of attractive interaction between the particles. We have observed that the binary system undegoes demixing transitions in the sticky sphere limit. Our simulations and analysis have provided insights into the conditions governing this first order phase transitions, with observable changes in particle configurations and thermodynamic properties. Through metrics such as the hexatic order parameter and radial distribution functions, we have quantified the extent of phase separation and elucidated the underlying structural transformations. The phase diagram constructed showcases distinct regions corresponding to mixed and phase-separated states, shedding light on the correlation between phase separation and orientational ordering. Our findings highlight the importance of range of intermolecular forces in dictating phase behaviour of binary mixtures of particles qith very short range attraction.

\begin{acknowledgments}
	The authors acknowledge financial support from the Department of Atomic Energy, India 
	through the plan project (RIN4001-SPS).  
\end{acknowledgments}

\bibliography{qsps} 
\end{document}